 \documentclass[12pt,preprint]{aastex}

\slugcomment{Submitted to P.A.S.P.}

\shorttitle{Elliminating the Coriolis effect}
\shortauthors{Hickson}

\begin{document}


\title{Elliminating the Coriolis Effect in Liquid Mirrors}


\author{P. Hickson}
\affil{Department of Physics \& Astronomy, University of British Columbia,
    6224 Agricultural Road, Vancouver, BC V6T1Z1, Canada}







\begin{abstract}
If uncorrected, the Coriolis force due to the rotation of the Earth causes
significant aberration of images produced by large liquid-mirror telescopes. 
We show that this problem can be eliminated by a 
fixed compensating tilt of the liquid-mirror rotation axis. The required
tilt angle, which is a function of latitude and mirror rotation rate,  is of order 
10 arcsec for current telescopes. This result removes
the last fundamental obstacle to achieving diffraction-limited performance
with large liquid mirrors.

\end{abstract}




\section{Introduction}

Liquid-mirror telescopes employ a rotating primary mirror surfaced
with a metallic liquid, usually mercury, to reflect and focus light 
\citep{b82,h.94,pm97}. Laboratory tests and astronomical observations
\citep{b.92, hm98, hm01} indicate that liquid mirrors can provide an 
optical surface that is parabolic to within a fraction of a wave
and provide astronomical-quality images. Current liquid-mirror
telescopes are zenith-pointing, however, their comparatively low 
cost makes them competitive with conventional telescopes for 
many types of observations such as wide-angle surveys.

Departures from the ideal parabolic shape, due to various influences
such as gravitational nonuniformity and Lunar tides, have been analysed 
by \cite{gh92} and \cite{m00}. These authors found that these
influences have a negligible effect on the images produced by 
liquid mirror telescopes with the exception of the Coriolis force that results
from the Earth's rotation. They found that, for an ideal non-viscous
fluid, the Coriolis effect introduces astigmatism and coma that results 
in a image spread that could be as large as $\sim 1.7$ arcsec for a 
2.7m f/1.5 telescope to  $\sim 6.7$ arcsec for a 10m f/1.5 telescope.
This is clearly a very significant effect, potentially being larger 
even than the atmospheric seeing.

While the Coriolis aberrations might in principle be removable by a 
suitably-designed optical corrector, the exact correction required
is difficult to determine because of hydrodynamic and viscous effects 
in the fluid layer that covers the mirror. The presence of any non-axisymmetric 
force will cause a periodic time-dependent flow of the fluid. In addition 
to greatly complicating the modeling of the system, such a flow will 
result in print-through of any imperfections in the surface of the 
structure that supports the fluid layer, making complete correction 
of the images very difficult.

In this paper it is shown that the Coriolis effect can in fact be
eliminated by means of a tilt of the rotation axis of
the mirror. When this is done, there are no significant time-dependent 
forces in the rotating frame of the mirror, and the fluid surface is static.
Because there is no fluid flow there are no hydrodynamic effects and
the mirror's surface is a paraboloid to within a small fraction
of the wavelength of light.

\section{Eliminating the Coriolis effect}

In the analysis that follows, we assume that sufficient time has
elapsed from the start of rotation to allow the fluid to come to
equilibrium. It is then sufficient to consider only inertial
and gravitational forces; other effects such as surface tension
have been shown by direct laboratory tests (\cite{b.92} to have 
a neglible effect on the shape of the surface. We will justify the
assumption of equilibrium by showing that, when the mirror rotation
axis is chosen appropriately, there are no significant time-dependent 
forces acting on the fluid that would cause a departure from equilibrium.

Consider a liquid mirror telescope located at latitude $l$, The axis
and rate of rotation of the liquid mirror is defined by
the angular velocity vector $\mathbf\Omega_0$. In the rotating
frame of the Earth, the instantaneous velocity of an element of fluid 
is given by
\begin{equation}
	\mathbf{v_0 = \Omega_0\times r},
\end{equation}
where 
$\mathbf r$ is the vector extending from the vertex of the mirror 
(the point where the rotation axis intersects the surface of the
fluid) to the fluid element.

Now consider an inertial frame moving, but not rotating, with the 
Earth. We can ignore the small acceleration due to the Earth's motion
around the Sun. The resulting centrifugal force is balanced by
the Sun's gravitational attraction, and it has already been shown
that the tidal force of the Moon, and therefore the smaller tidal
force of the Sun, are unimportant \cite[]{gh92}. In the
inertial frame, the fluid is also rotating about the Earth's axis
with velocity
\begin{equation}
	\mathbf{v = \Omega_\earth\times(R_\earth+r)},
\end{equation}
where $\mathbf\Omega_\earth$ is
the angular velocity vector of the rotating Earth and $\mathbf R_\earth$
is the vector extending from the center of the Earth to the
vertex of the mirror. The velocity of the fluid element in the
inertial frame is thus
\begin{eqnarray}
	\mathbf{v_e} & = & \mathbf{v_0 + v}  \nonumber \\
	& = & \mathbf{\Omega_e \times r + \Omega_\earth\times R_\earth}, \label{eqv}
\end{eqnarray}
where
\begin{equation}
	\mathbf{\Omega_e = \Omega_0 + \Omega_\earth}
\end{equation}
is the effective angular velocity vector.

Eqn (\ref{eqv}) shows that the effect of the Earth's rotation is twofold.
First, the effective axis and rate of rotation is changed to $\mathbf\Omega$, 
the vector sum of $\mathbf\Omega_0$ and $\mathbf\Omega_\earth$. Second, an 
acceleration 
\begin{eqnarray}
	\mathbf {a_C} & = & {d\mathbf V\over dt} \nonumber \\
	& = & \mathbf{\Omega_\earth\times(\Omega_\earth\times R_\earth)}
\end{eqnarray}
occurs, directed inward toward the Earth's axis, with magnitude
\begin{equation}
	a_C = R_\earth\Omega_\earth^2 \cos l. \label{eqac}
\end{equation}

Consider now the gravitational force acting on the fluid. The gravitational acceleration
is
\begin{eqnarray}
	\mathbf g & = & -{GM_\earth\over \left |\mathbf{R_\earth+r}\right |^3}
		(\mathbf{R_\earth+r}) \nonumber \\
	& = & \mathbf{g_0 + g_T}, \label{eqg}
\end{eqnarray}
where 
\begin{equation}
	\mathbf g_0 = -{GM_\earth\over R_\earth^3}\mathbf R_\earth
\end{equation}
is the gravitational acceleration at the mirror's vertex and
\begin{equation}
	\mathbf g_T = -{GM_\earth\over \left |\mathbf{R_\earth+
		r}\right |^3}\mathbf{(R_\earth+r) - g_0} \label{eqgt}
\end{equation}
is a small tidal term representing the differential acceleration across the mirror due to
the gradient and divergence of the gravitational field.

The total centripetal acceleration of the fluid element must equal the gravitational
acceleration, which is the only significant external force on the fluid. Therefore,
\begin{equation}
	\mathbf{\Omega_e\times(\Omega_e\times r) + a_C = g_0 + g_T} \label{eqa}
\end{equation}

Ignoring for the moment the small tidal term $\mathbf{g_T}$, to which we shall
return later, it is now evident that if the total angular velocity vector 
$\mathbf\Omega$ is parallel to the effective gravitational acceleration 
$\mathbf{g_e = g_0-a_C}$, the 
force acting on the fluid will be symmetric about the axis defined by 
$\mathbf\Omega$. To see this observe that, when the fluid is in equilibrium, the
radius $r$ of any fluid element is constant, so the centripetal acceleration 
$\mathbf{\Omega_e\times(\Omega_e\times r)}$ has constant magnitude and is
directed toward the rotation axis. By definition, the effective gravitational acceleration
$\mathbf{g_e}$ acts along the rotation axis, so the fluid element feels only
constant radial and axial forces, and no azimuthal forces. Because these forces
are constant, there will be no further flow of the fluid once it has reached equilibrium
and therefore no hydrodynamic effects. The required geometrical condition is 
illustrated in Fig. 1. Since we are free to choose the orientation
of the mirror rotation axis, it is always possible to achieve this condition. 
{\it This shows that the Coriolis effect can be
eliminated by tilting the mirror rotation axis in order to make the effective angular
velocity vector $\mathbf{\Omega_0 + \Omega_\earth}$ parallel to the effective
acceleration vector $\mathbf{g_0-a_C}$}.

Referring to Fig. 1, we see that the required tilt angle, with respect to the
zenth, is $\alpha - \beta$ where
\begin{eqnarray}
	\alpha & = & {\mathbf{g_e\cdot g_0}\over g_e g_0} \nonumber \\
	& \approx & {\Omega_\earth^2R_\earth\sin l\over g_0}  \label{eqalpha}
\end{eqnarray}
and
\begin{eqnarray}
	\beta & = & {\mathbf{\Omega_e\cdot\Omega_0}\over\Omega_e\Omega_0} \nonumber \\
	& \approx & {\Omega_\earth\cos l\over\Omega_0} \label{eqbeta}
\end{eqnarray}

In practice, liquid mirrors are leveled using a precise bubble level which is
of course sensitive to the effective vertical direction defined by $\mathbf{g_e}$,
not the true vertical. In this case, the required tilt angle is just $\beta$. From
Fig. 1. it is evident that the required axis tilt is in the North-South direction and
towards the equator if the mirror rotates in the same sense as the Earth, 
(counterclockwise in the northern hemisphere and clockwise in the southern 
hemisphere) and away from the equator otherwise.

Now, $\Omega_\earth \approx 7.29\times 10^{-5}$ s$^{-1}$, $R_\earth\approx 6.38\times 10^6$ m,
$g_0 \approx 9.81$ ms$^{-2}$ and $\Omega_0^2 \approx (g_0/2F)$, where $F$ is
the telescope focal length. Substituting these quantities in Eqns 
(\ref{eqalpha}) and (\ref{eqbeta}) gives
\begin{eqnarray}
	\alpha & \approx & 3.46\times 10^{-3} \sin l  \label{eqalpha2} \\
	\beta & \approx & 3.29\times 10^{-5} F^{1/2} \cos l \label{eqbeta2}
\end{eqnarray}
where $F$ is measured in meters.

\section{The equilibrium surface and the effect of the tidal field}

Let us now assume that the Coriolis-cancelling condition described in
the previous section is
satisfied. What is the shape of the mirror's surface? To calculate this,
it is convenient to use cylindrical coordinates ($\rho, \phi,z$) with the
$z$ axis aligned with the effective rotation axis $\mathbf\Omega_e$, and
origin at the vertex of the mirror. The orientation is chosen so that $z$
increases in the outward direction (ie. away from the Earth) and 
$\phi = 0$ is the North direction.

Rather than work with the vector equation (\ref{eqa}), we introduce the
scalar potential $\Phi$, whose gradient is the acceleration. The
total potential of a fluid element is then
\begin{equation}
	\Phi = -{1\over 2}\Omega_e^2 \rho^2 + g_ez + \Phi_T. \label{eqphi}
\end{equation}
In this equation, the first term on the right hand side is the centrifugal
potential arising from rotation about the axis $\mathbf\Omega_e$, 
the second term represents the effective gravitational acceleration and
the third term is the potential corresponding to the tidal term
$\mathbf{g_T}$. The zero point of the potential is arbitrary because any
constant can be added without affecting the acceleration. We have
chosen $\Phi = 0$ at the mirror vertex $(\rho = mz = 0)$.

In equilibrium, the surface of a liquid has constant potential, so Eqn (\ref{eqphi})
gives the surface shape directly,
\begin{equation}
	z = {\Omega_e^2 \over 2g_e}\rho^2 - {1\over g_e}\Phi_T \label{eqz}
\end{equation}
which is a paraboloid of focal length
\begin{equation}
	F = {1\over 4}\left ({dz\over d\rho^2}\right )^{-1} = {g_e\over 2\Omega_e^2} \label{eqf}
\end{equation}
with a small correction due to the tidal term. 
The quantities $\Omega_e$
and $g_e$, the effective angular velocity and gravitational acceleration,
can be determined by applying the cosine rule to the geometry of Fig 1.
and using Eqn (\ref{eqac}). The result is
\begin{eqnarray}
	\Omega_e & = & \left [ \Omega_0^2+\Omega_\earth^2+2\Omega_0
		\Omega_\earth\sin(l-\epsilon) \right ]^{1/2} \\
	g_e & = & \left [ g_0^2+(\Omega_\earth^4R_\earth^2-
		2g_0\Omega_\earth^2R_\earth)\cos^2l \right ]^{1/2}
\end{eqnarray}

Let us now determine the effect of the tidal term. Eqn (\ref{eqgt}) 
shows that the tidal term represents the difference between the
actual gravitational acceleration and a hypothetical uniform gravitational
acceleration equal to the value at the mirror vertex.  However, because
of the Earth's rotation,  the effective axis of rotation of the mirror does 
not pass through the centre of the Earth, but is instead tilted from the 
vertical. Thus, there is a tilt between our cylindrical coordinate system 
and a geocentric system. To first order in the tilt angle $\alpha$, the tidal 
potential is therefore
\begin{equation}
	\Phi_T = -{GM_\earth\over \left |\mathbf{R_\earth+r}\right |}
		-{GM_\earth\over R_\earth^2}(z-\alpha\rho\cos\phi)
		+{GM_\earth\over R_\earth} \label{eqphit}
\end{equation}
where the last term is a constant chosen to make $\Phi_T = 0$ at the
mirror vertex. To first order in $\alpha$, we have
\begin{equation}
	\left |\mathbf{R_\earth+r}\right | = \left [(R_\earth+z-
	\alpha\rho\cos\phi)^2+\rho^2 \right ]^{1/2} \label{eqr}
\end{equation}
Using Eqn (\ref{eqr}), the first term on the right-hand side of Eqn (\ref{eqphit})
can be expanded in a power series in $\rho/R_\earth$ and $z/R_\earth$ and
$z$ can be eliminated by substituting Eqn (\ref{eqz}). From Eqn (\ref{eqz}),
dividing this result by $-g_e$ then gives the resulting surface deformation $z_T$.
To fourth order in $\rho$ and first order in $\alpha$, the result is
\begin{equation}
	z_T = {g_0\over g_e}\left\{ {\rho^2\over 2R_\earth}
		-\left  [6+6{R_\earth\over F} +{R_\earth^2\over F^2} \right ]
		{\rho^4\over 16R_\earth^3}-\left [ 3+{R_\earth\over F}\right ]
		{\alpha\rho^3\cos\phi\over 2R_\earth^2}\right\} \label{eqzt}
\end{equation}
The first two terms in this equation are already known and correspond to a 
small shift in focal lenth and a small amount of 
spherical aberration. The new
focal length (from Eqns \ref{eqf}, \ref{eqz} and \ref{eqzt}) is
\begin{equation}
	F_e = F\left [1+{2F\over R_\earth}\right ]^{-1}.
\end{equation}
The maximum amplitude of the second term is approximately 
$D^4/256 F^2R_\earth$, where $D$ is the diameter of the mirror. 
For a 10m f/2.5 mirror it is $10^{-7}$ m. At a wavelength of 1 um, this 
corresponds to a fifth of a wave of spherical aberration, which can 
easily be corrected optically if necessary.

The third term is a very small comatic distortion that arises from the misalignment 
of the tidal field with the effective rotation axis. The maximum amplitude 
of this term is $\alpha D^3/16 F R_\earth$. For a 10-m f/2.5 mirror located
at 30 degrees latitude it has the value $6.8\times 10^{-10}$ m, which is
less than a thousandth of a wave. Even for a 100-m telescope the maximum
error is less than a tenth of a wave. It follows that this term can be safely
ignored for all liquid-mirror telescopes currently conceivable.

\section{Discussion}

Until now, the Coriolis effect was considered to be a potentially signigicant
problem for large liquid-mirror telescopes. While viscous effects might 
reduce the the surface distortion to some degree, the predicted effect is so
large for 10-m class mirrors that it was a cause of concern.

In this paper we have shown that the Coriolis effect can in fact be eliminated by 
a fixed compensating tilt of the axis of rotation of the liquid mirror. A small 
residual tidal effect was shown 
to be of no significance to all present and forseen liquid-mirror telescopes.
This result removes the last fundamental obstacle to achieving diffraction-limited
performance with large liquid mirrors.

In order to eliminate the Coriolis effect, the axis of rotation of the mirror
must be tilted from the effective vertical (defined by a pendulum or bubble
level) by the angle $\beta$ given by Eqns (\ref{eqbeta}) and (\ref{eqbeta2}).
For the 3-m f/1.5 mirror of the Nasa Orbital Debris Observatory \cite[]{pm97}, 
which rotates counterclockwise, the required shift is to the South by an angle of 12.1
arcsec. Axis tilts of this magnitude have been made with this telescope, in
order to investigate the effects on the image \cite[]{m00}, but the effects
were  smaller than the atmospheric seeing and could not be adequately assessed.
For the 6-m f/1.5 mirror of the Large Zenith Telescope \cite[]{h.98}, 
the tilt required is 13.1 arcsec. The larger mirror diameter and better image
sampling should allow us to verify the technique using this 
telescope.






\acknowledgments

I am grateful to the National Solar Observatory and NASA for hospitality during 
visits to Sunspot and the NASA Orbital Debris Observatory, where part of this work 
was conducted. I thank Mark Mulrooney for numerous discussions. This work was 
supported by grants from the Natural Sciences and Engineering Research Council 
of Canada.




\clearpage


\begin{figure}
\plotone{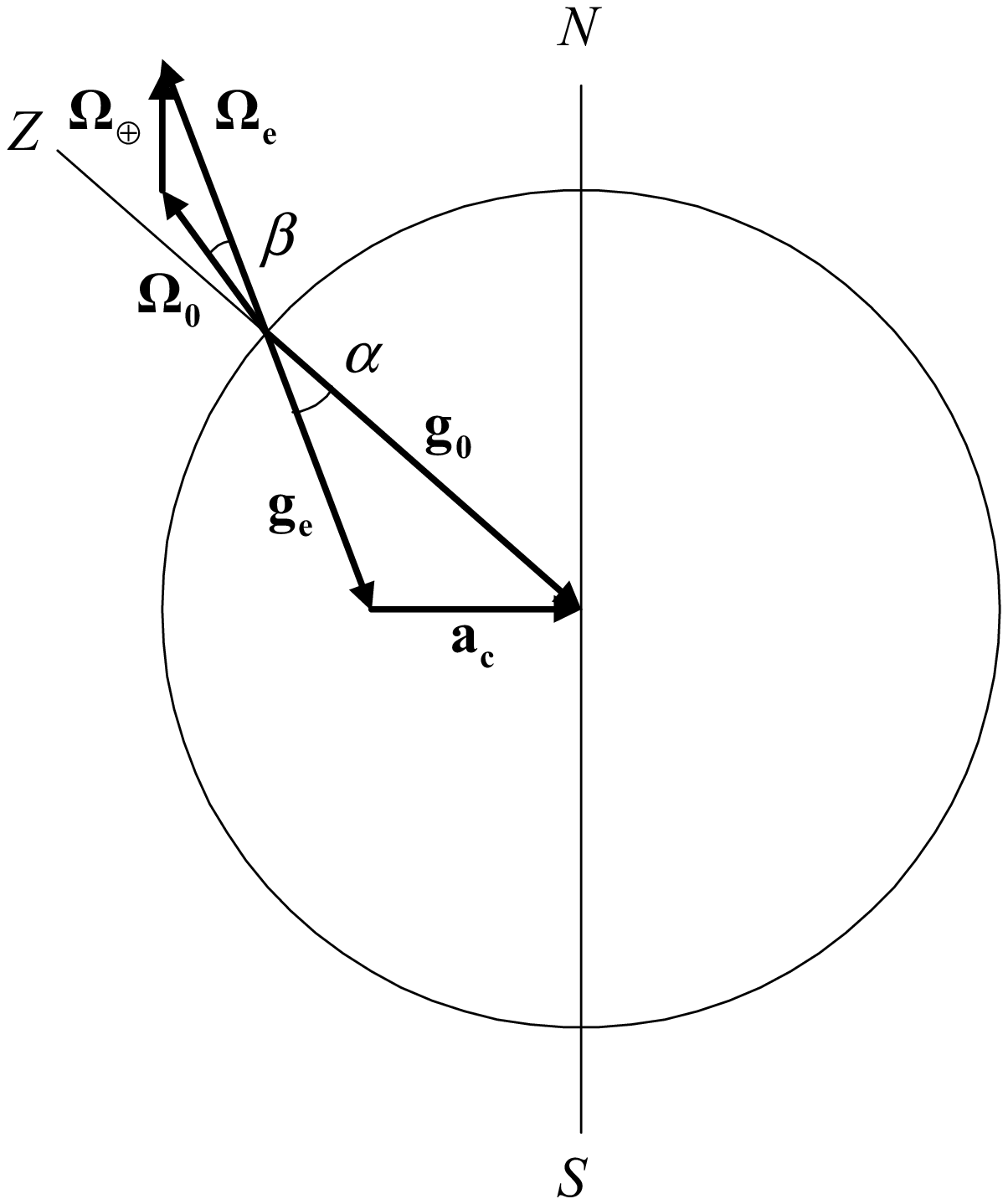}
\vspace{-8cm}
\caption{Geometry of angular velocity and acceleration vectors. The
circle represents the Earth, rotating about the N-S axis with angular
velocity $\mathbf{\Omega_\earth}$. The vertex
of the liquid mirror is located at the intersection of the circle and
the straight line extending to the zenith Z. In the rotating frame of
the Earth, the mirror rotates about axis $\mathbf{\Omega_0}$ but
in an inertial frame it rotates about the resultant axis 
$\mathbf{\Omega_e}$. To eliminate the Coriolis effect, this axis
must be parallel to the effective gravitational acceleration $\mathbf{g_e}$
\label{fig1}}
\end{figure}

\end{document}